# Thermophotovoltaic Efficiency of 40%

Alina LaPotin,[1] Kevin L. Schulte,[2] Myles A. Steiner,[2] Kyle Buznitsky,[1] Colin C. Kelsall,[1] Daniel J. Friedman,[2] Eric J. Tervo,[2] Ryan M. France,[2] Michelle R. Young,[2] Andrew Rohskopf,[1] Shomik Verma,[1] Evelyn N. Wang,[1] Asegun Henry[1a]

[1]Department of Mechanical Engineering, Massachusetts Institute of Technology, Cambridge MA, 02139, USA

[2]National Renewable Energy Laboratory, Golden CO, 80401, USA

[a]Author to whom correspondence should be addressed: ase@mit.edu

**Abstract**

**We report the fabrication and measurement of thermophotovoltaic (TPV) cells with efficiencies of >40%, which is a record high TPV efficiency and the first experimental demonstration of the efficiency of high-bandgap tandem TPV cells. TPV efficiency was determined by simultaneous measurement of electric power output and heat dissipation from the device via calorimetry. The TPV cells are two-junction devices comprising high-quality III-V materials with band gaps between 1.0 and 1.4 eV that are optimized for high emitter temperatures of 1900-2400°C. The cells exploit the concept of band-edge spectral filtering to obtain high efficiency, using high-reflectivity back surface reflectors to reject unusable sub-bandgap radiation back to the emitter. A 1.4/1.2 eV device reached a maximum efficiency of (41.1 ± 1)% operating at a power density of 2.39 W/cm$^2$ under an irradiance of 30.4 W/cm$^2$ and emitter temperature of 2400°C. A 1.2/1.0 device reached a maximum efficiency of (39.3 ± 1)% operating at a power density of 1.8 W/cm$^2$ under an irradiance of 20.1 W/cm$^2$ and emitter temperature of 2127°C. These cells can be integrated into a TPV system for thermal energy grid storage (TEGS) to enable dispatchable renewable energy. These new TPV cells enable a pathway for TEGS to reach sufficiently high efficiency and sufficiently low cost to enable full decarbonization of the grid. Furthermore, the high demonstrated efficiency also gives TPV the potential to compete with turbine-based heat engines for large-scale power production with respect to both cost and performance, thereby enabling possible usage in natural gas or hydrogen-fueled electricity production.**

**Main**

Thermodynamics initially developed as a framework for understanding and improving the performance of heat engines as they fueled the industrial revolution. Since then, heat engines have gone on to play a pivotal role in modern society, as more than 90% of electricity today is generated by heat engines. In reflecting on this history, it is remarkable that one specific class of heat engines emerged as the primary workhorse of the power industry, namely those using a turbine, which is used for electricity production from coal, natural gas, nuclear and concentrated solar. Turbines proliferated because of their high efficiency (25-60%) and their low cost per unit power (CPP) generated ($0.5-1/W). However, since turbines intrinsically require moving parts, there are corresponding requirements on the high-temperature mechanical properties of the materials of construction, since they are subject to centrifugal loads. Thus, they have reached their practical

limits in terms of cost and efficiency, barring a materials discovery that would allow them to operate at substantially higher turbine inlet temperatures than the current values of ~1500 °C for Brayton cycles and ~700 °C for Rankine cycles.[1]

Henry and Prasher[1] highlighted that solid-state heat engines, having no moving parts, possess an advantage in this sense, enabling their operation at significantly higher temperatures than turbines. Thermal energy grid storage (TEGS), as described by Amy et al.,[2] relies on thermophotovoltaics (TPV), a solid-state heat engine, to convert heat to electricity above 2000°C which is a regime inaccessible to turbines (see Fig. 1a). Although TEGS was initially conceived with a molten silicon storage medium in mind, a graphite storage medium is even lower cost ($0.5/kg), and the projected capital cost per unit energy (CPE) is less than $10/kWh.[3] This cost is so low, it would allow TEGS to meet the proposed cost targets (< $20/kWh) for long duration energy storage that would allow renewable energy with storage to be cost-competitive with fossil fuels.[4-6] As a result, the proliferation of TEGS could ultimately enable abatement of ~40% of global $CO_2$ emissions, by decarbonizing the grid (~25% of emissions), and then enabling carbon-free electricity to charge vehicles in the transportation sector (~15% of emissions).[7]

The work of Amy et al.[2,8] showed initial feasibility for operating a thermal energy storage infrastructure above 2000°C, but the linchpin to realizing the full system is the demonstration of high efficiency TPV cells. Like photovoltaics (PV), TPV uses the photoelectric effect to convert light to electricity. However, with TPV, the source of light is a terrestrial object that is heated to a high temperature that predominantly emits infrared (IR) light to the TPV cell, as opposed to light from the sun. In 1980 Swanson showed that a silicon TPV cell with an integrated back surface reflector (BSR) and a tungsten emitter at 2000°C could reach 29% efficiency.[9] In the ~40 years since Swanson's initial demonstration (see Fig. 1b), TPV fabrication and performance have greatly improved using III-V semiconductors and the development of high-quality BSRs.[10,11] These results have opened the path to achieving much higher efficiency by replicating similar advancements with materials that have higher bandgaps, multiple junctions, and using higher emitter temperatures. However, despite the fact that a number of studies have presented model predictions that TPV efficiencies can exceed 50%,[10,12,13] the demonstrated efficiencies are still only as high as 32%, but remarkably at much lower temperatures < 1300°C.[12-14]

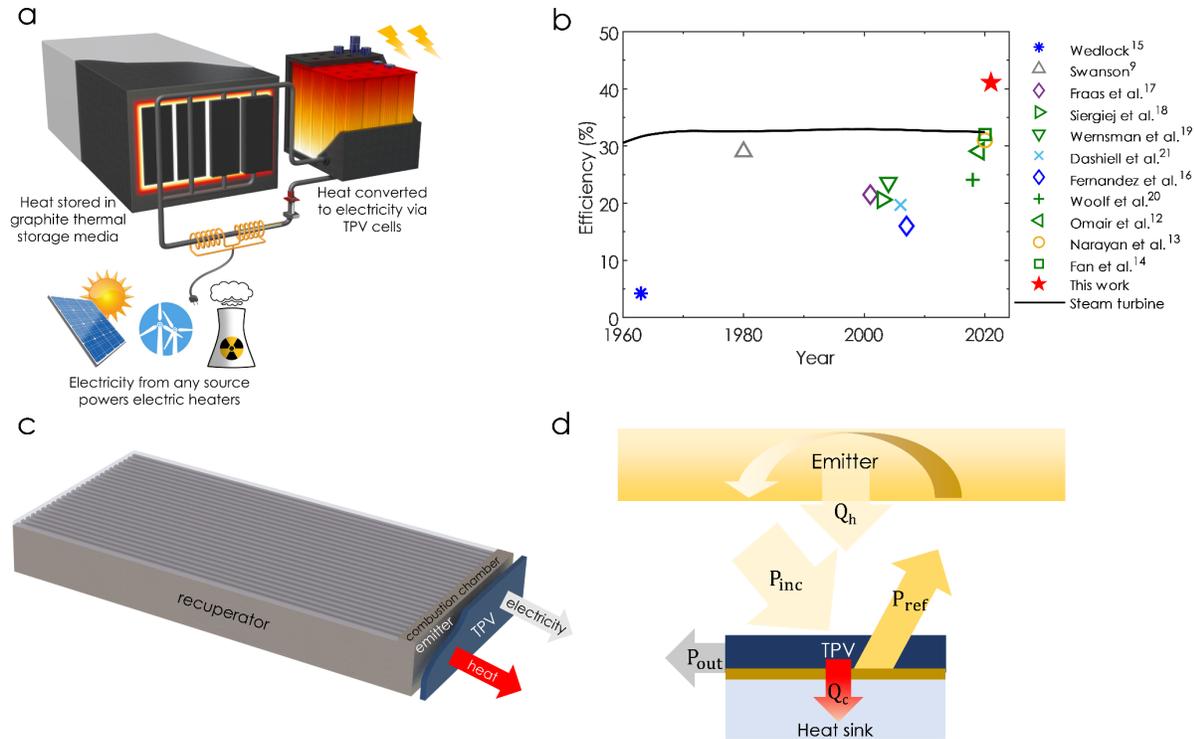

**Figure 1.** a) Conceptual illustration of TEGS, which uses TPV for conversion of heat to electricity. TEGS takes in electricity, converts it to heat via joule heating, which is then transferred using liquid metal tin to a storage bank of graphite blocks. The blocks store the heat, and then when electricity is desired, the liquid metal retrieves the heat and delivers it to a power block containing TPV cells that convert light emitted by the hot infrastructure. b) Historic TPV efficiency[11] with cell materials Ge[15,16] (blue), Si[9] (gray), GaSb[17] (purple), InGaAs[12,14,18-20] (green), InGaAsSb[21] (light blue), and GaAs[13] (yellow). The black line shows the average thermal efficiency of power generation in the U.S. using a steam turbine (coal and nuclear).[22,23] Prior to 2000, turbine efficiencies shown also include natural gas. c) Conceptual illustration of a combustion-based electricity generation system using TPV. The system consists of an all ceramic recuperator, similar to a printed circuit heat exchanger, with the end comprising a combustion chamber. Air is preheated by exhaust and then combined with fuel for combustion near the end facing the TPV. The hot exhaust then delivers heat to the ceramic which radiates it to the TPV. c) Energy inputs and outputs of a TPV system.

Here we report TPV efficiency measurements of > 40%, measured directly through calorimetry. This is a record high TPV efficiency and the first experimental demonstration of high-bandgap tandem TPV cells. This is also the first demonstration of a solid-state heat engine (terrestrial heat source) with efficiency higher than the average heat engine efficiency in the U.S., which is < 35% based on primary energy inputs and electricity output.[24] An efficiency of 40% is also higher than most steam cycles, and is in the same range as simple cycle gas turbines.[25] Thus, 40% represents a major step forward (see Fig. 1b), as this is the first time another type of heat engine has the potential to compete with turbines by exhibiting comparable efficiency and potentially even lower CPP e.g., < $0.25/W.[2,26]

To properly contextualize why this has broad reaching implications, it should be appreciated that over the last century, a range of alternative heat engines such as thermoelectrics,[27] thermionics,[28] TPV,[11] thermally regenerative electrochemical systems,[29] thermoacoustic engines,[30] and Stirling engines[31,32] have been developed. All these technologies have some intrinsic advantage(s) over turbines, such as low maintenance, no moving parts, and/or easier integration with an external heat source, yet none of them can compete with the efficiency and CPP of turbines for large-scale heat to electricity conversion. The temperature regime examined here is accessible via combustion of natural gas or hydrogen, which could be made into an efficient power generation system by using recuperators made from refractory metals and oxides (for example Fig. 1c).[17,33] With TPV now being a potentially cheaper alternative to a turbine with comparable performance, a broad range of potential applications, such as TEGS,[2,34] natural gas or hydrogen-fueled power generation[17,33,35-40] and high-temperature industrial waste heat recovery are now feasible.

**High Efficiency TPV Cells**

It is important to appreciate that the efficiency of a TPV cell is defined differently from that of a solar cell because, unlike a solar cell, a TPV system can preserve and later convert the energy associated with sub-bandgap photons. This is because in the contexts where TPV is envisioned to be used, the TPV cell has a high view factor to the emitter, meaning that sub-bandgap photons can be reflected back to the emitter by the TPV cell (Fig. 1d), in contrast to the situation of a solar cell and the sun. By reflecting unconverted photons, the energy of the sub-bandgap light is preserved through reabsorption by the emitter. The reflected light helps to keep the emitter hot, thereby minimizing the energy input required to heat the emitter. As a result, the efficiency of a TPV cell is given by,

$$\eta_{TPV} = \frac{P_{out}}{P_{out}+Q_c} = \frac{P_{out}}{P_{inc}-P_{ref}} \qquad (1)$$

In Eq. 1, $P_{out}$ is the electric power generated by the TPV cell (i.e., $P_{out} = V_{oc}I_{sc}FF$), where $V_{oc}$ is the open circuit voltage, $I_{sc}$ is the short circuit current and $FF$ is the fill factor of the IV curve. The total heat absorbed and generated in the cell is denoted by $Q_c$, which is comprised of the heat generated by parasitic absorption in the semiconductor or metal reflector, thermalization losses due to excess incident photon energy, Joule heating losses due to current flow, and nonradiative recombination losses. The net energy received by the cell is equivalent to $P_{out} + Q_c$ and can also be expressed as $P_{inc} - P_{ref}$ where $P_{inc}$ is the incident energy and $P_{ref}$ is the reflected energy. Based on Eq. 1, to increase TPV efficiency, one must increase the power output $P_{out}$ or reduce the amount of heat absorbed and generated in the cell ($Q_c$).

The TPV system efficiency is defined as $\eta_{TPV,system} = P_{out}/Q_h$, where $Q_h$ is the energy leaving the emitter. A system's efficiency may be less than the TPV efficiency due to view factor related losses or other heat losses from the emitter to the environment. However, $\eta_{TPV} = \eta_{TPV,system}$ if these losses are mitigated, which can be the case for TEGS, or a large scale combustion system, in which case $Q_h = P_{out} + Q_c$.[2,26] This can be accomplished by increasing the scale of the system such that the heated material has a large volume to surface area ratio and heat losses from the surfaces can be minimized with proper insulation,[26] and if the emitter surface and TPV module have a large surface area to perimeter ratio such that the view factor between them approaches one. This is a critically important aspect of achieving a high value for $\eta_{TPV,system}$.[2,26] Considering that the other losses can be negligible, $\eta_{TPV}$ is the metric we use here because it is a conventional and generalizable metric used to describe the performance of a cell/emitter pair independent of system-level characteristics.[11]

The high emitter temperatures targeted here for TEGS and other applications allow higher bandgap cells ≥1.0 eV to be used instead of the low bandgap, InGaAs- or GaSb-based cells traditionally used for TPV. This is key, because the spectrum of light redshifts toward longer wavelengths as the radiator temperature is lowered, which is why traditional TPV cells that are paired with <1300°C emitters are typically based on ~0.74 eV InGaAs or ~0.73 GaSb. Considerable work on low bandgap semiconductors has been undertaken with the envisioned application of converting heat from natural gas combustion,[17,33,35-40] concentrated solar power,[26] space power applications,[41,42] and more recently energy storage.[2,34,43] This pioneering body of work has led to the identification of three key features that now enable TPV to become a competitive option for converting heat to electricity commercially, namely: (1) the usage of higher bandgap materials in combination with higher emitter temperatures, (2) high-performance multi-junction architectures with bandgap tunability enabled by high-quality metamorphic epitaxy,[44] and (3) the integration of a high reflectivity BSR for band-edge filtering.[10,12]

With respect to higher bandgaps, they increase efficiency because there is an almost constant penalty on voltage of ~0.3-0.4 V, due to the thermodynamic requirements on the radiative recombination rate.[45] As a result, this unavoidable loss penalizes lower bandgap cells more than higher bandgap cells, because this loss comprises a smaller fraction of the voltage for higher bandgap materials. Using higher bandgap materials also needs to be accompanied by operation at higher temperatures to maintain sufficiently high power density, which scales with the emitter temperature to the fourth power. Operation at high power density is critical for TPV economics because the cell costs scale with their area, and if the power generation per unit area increases the corresponding CPP decreases.[1]

With respect to BSRs, a highly reflective BSR is critical to minimize $Q_c$. Highly reflective BSRs provide the additional benefit of boosting open-circuit voltage, because they also improve recycling of luminescent photons generated via radiative recombination.[46-48] This effect has led to regular integration of BSRs with solar PV cells, which provides a template for their use in TPVs. With these important lessons from prior work in mind, the cells developed here are 1.2/1.0-eV and 1.4/1.2-eV two-junction designs intended for the TEGS application with emitter temperatures between 1900-2400°C.[2] Multi-junction cells increase efficiency over single junctions by reducing hot carrier thermalization losses and reducing resistive losses by operating at a lower current density. The cells were based on the inverted metamorphic multi-junction (IMM) architecture pioneered at NREL.[49-51]

The first cell design uses lattice-mismatched 1.2 eV AlGaInAs and 1.0 eV GaInAs top and bottom junctions, where the lattice mismatch is with respect to the crystallographic lattice constant of the GaAs substrate on which they are grown. The second design uses a lattice-matched 1.4 eV GaAs top cell and a lattice-mismatched 1.2 eV GaInAs bottom cell, taking advantage of the inherently higher material quality of lattice-matched epitaxy in the GaAs cell. The lower bandgap 1.2/1.0-eV tandem offers the potential for higher power density than the 1.4/1.2-eV tandem because it converts a broader band of the incident spectrum, and consequently the requirements on the BSR are less stringent to obtain high efficiency.[43] Higher power density can also be a practical engineering advantage. On the other hand, while the 1.4/1.2-eV tandem has a lower power output, the reduced current density of this bandgap combination potentially enables higher efficiency than the 1.21.0-eV tandem if resistive losses are an issue.

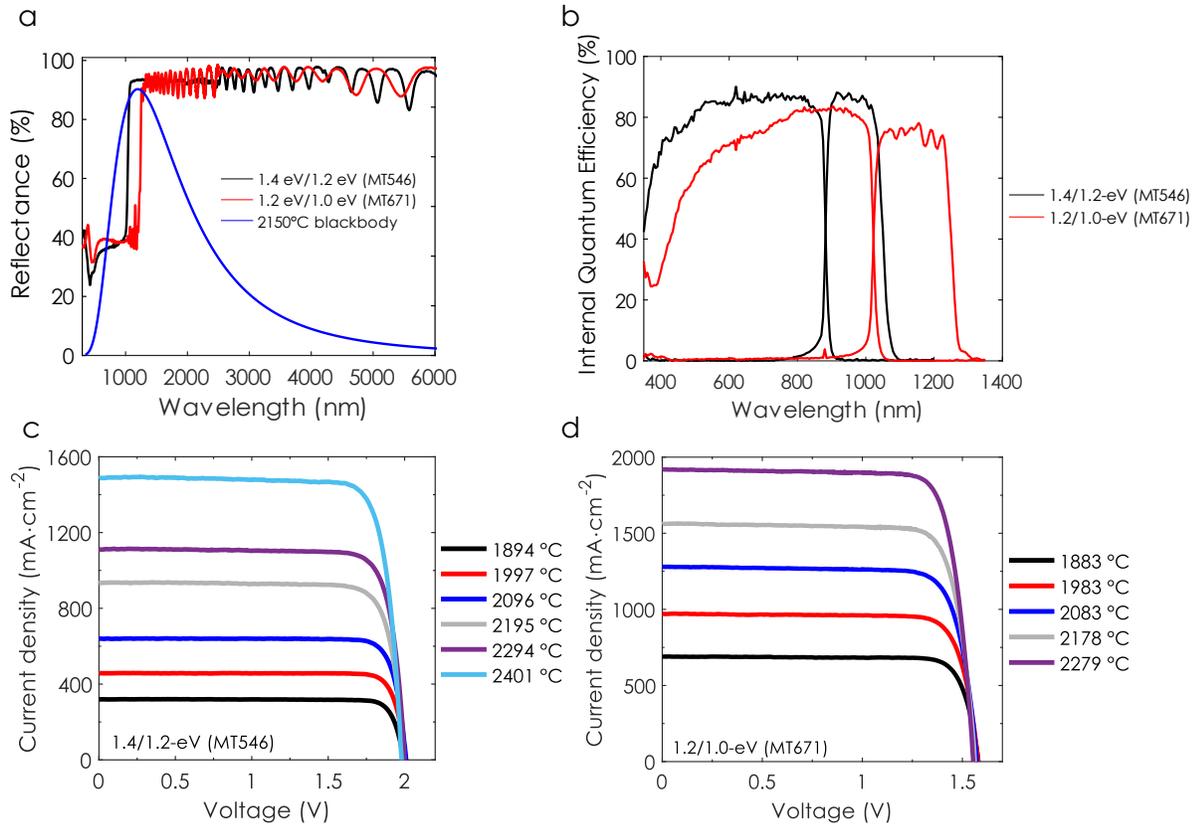

**Figure 2**. a) Reflectance of the 1.4/1.2-eV and 1.2/1.0-eV tandems. The 2150 °C blackbody spectrum is shown for reference. 2150 °C is the average emitter temperature in the TEGS operating range. b) Internal quantum efficiency (IQE) of the 1.4/1.2-eV and 1.2/1.0-eV tandems. The EQE is shown in Extended Data Fig. 1. c) Current density/voltage curves measured in the efficiency setup at varying emitter temperatures for 1.4/1.2-eV and d) 1.2/1.0-eV tandems.

**TPV Efficiency Measurement Results**

The TPV cell fabrication, measurement, and modeling details are provided in the Methods section. The cell labeled MT546 is a 1.4/1.2-eV tandem and that labeled MT671 is a 1.2/1.0-eV tandem. Reflectance measurements are shown in Fig. 2a and internal quantum efficiency (IQE) is given in Fig. 2b. The sub-bandgap spectral weighted reflectance for the 2150°C blackbody spectrum is 93.0% for MT546 and 93.1% for MT671. Current density vs voltage measurements were performed under a tungsten halogen bulb emitter and results for a range of emitter temperatures relevant to the TEGS application (~1900 °C – 2400 °C) are shown in Fig. 2c and Fig. 2d. As expected, the 1.2/1.0-eV tandem shows lower voltage but higher current density than the 1.4/1.2-eV tandem. The non-monotonic change in $V_{oc}$ at the highest emitter temperatures is because of increasing cell temperature (Extended Data Fig. 2) due to the presence of a heat flux sensor (HFS) used for the efficiency measurement, that undesirably also impedes heat flow. Fig. 3a shows the efficiency measurement at the same range of emitter temperatures and was accomplished by simultaneously measuring $Q_c$ and $P_{out}$.

The results for the higher bandgap tandem, MT546, show increasing efficiency with increasing emitter temperature, and the efficiency exceeds 40% at 2350 °C, which is within the target range of 1900-2400°C needed for the TEGS application. At 2400 °C, the efficiency is as high as 41.1 ±1%, while the average efficiency between 1900-2400 °C is 36.2%. The electrical power density is 2.39 W/cm² at the maximum emitter temperature of 2400°C. The rate of increase of efficiency with temperature slows at high emitter temperatures due to a reduction in FF due to increasing series resistance losses and the diminishing increase in $J_{sc}$ due to the cell becoming current-limited by the bottom cell at ~2250 °C.

The results for the lower bandgap tandem, MT671, show greater efficiency than MT546 at lower emitter temperatures due to the lower bandgaps of MT671. The efficiency of MT671 reaches a maximum of 39.3 ±1% at 2127 °C, quite close to 2150 °C which is the temperature at which our device model predicted this bandgap combination to be optimal.[43] The average efficiency between 1900-2300 °C is 38.2% and the efficiency remains high across a 400°C range of emitter temperatures. This is particularly noteworthy for the TEGS application because it suggests consistently high efficiency can be achieved even as the emitter temperature varies during the discharging process of the TEGS system. The reduction in efficiency beyond this temperature is due to the increasing series resistance losses and the diminishing increase in $J_{sc}$ due to the cell becoming current-limited by the bottom cell at temperatures greater than 2150 °C. The electrical power density is 2.42 W/cm² at the maximum emitter temperature measured of 2279 °C, and it is 1.81 W/cm² at the maximum efficiency point at the emitter temperature of 2127°C. Comparing the performance of the two cells across the range of emitter temperatures, they exhibit different characteristics which are advantageous for TEGS. The efficiency of MT671 is less sensitive to changes in emitter temperature, has a higher electrical power density at a given emitter temperature, and has a higher efficiency averaged over the emitter temperatures. However, MT546 can reach higher efficiency at the highest emitter temperatures.

Fig. 3a also shows model predictions for efficiency and corresponding uncertainty of the model prediction (see methods for a detailed explanation of this calculation). The good agreement obtained between the modeled and measured performance supports and validates the accuracy of the efficiency measurement and of the calorimetry-based method used to measure efficiency. Additionally, the good agreement suggests that the model can be extended to extrapolate how the performance would change with additional improvements, or at other operating conditions. The most important TPV cell property that could be improved is its spectral-weighted sub-bandgap reflectance, $R_{sub}$. Fig. 3b shows how the efficiency would change if $R_{sub}$ could be increased. To extrapolate the results to a real TPV system, here we assume that the emitter is tungsten (W), as it is in the TEGS system, and that the area ratio between the emitter and cell is $AR=1$, the view factor is $VF = 1$, and the cell temperature is 25 °C. A comparison between this spectrum and the spectrum under which the cells were measured (spectrum measurements shown in Extended Data Figure 3) can be found in Extended Data Figure 4. In this prediction, for a 2200 °C emitter temperature, the efficiency of MT546 exceeds 50% at $R_{sub} = 97\%$. The reason this is noteworthy is because the present value of $R_{sub}$ is considerably lower than what was achieved with the air bridge approach recently demonstrated by Fan et al.[14] Their work demonstrating a reflectivity > 98% charts a pathway towards further efficiency improvements. If the air bridge approach developed by Fan et al. could be combined with the advancements demonstrated here, it could lead to efficiencies greater than 56% at 2250°C, or greater than 51% averaged over the 1900-2400°C temperature range.

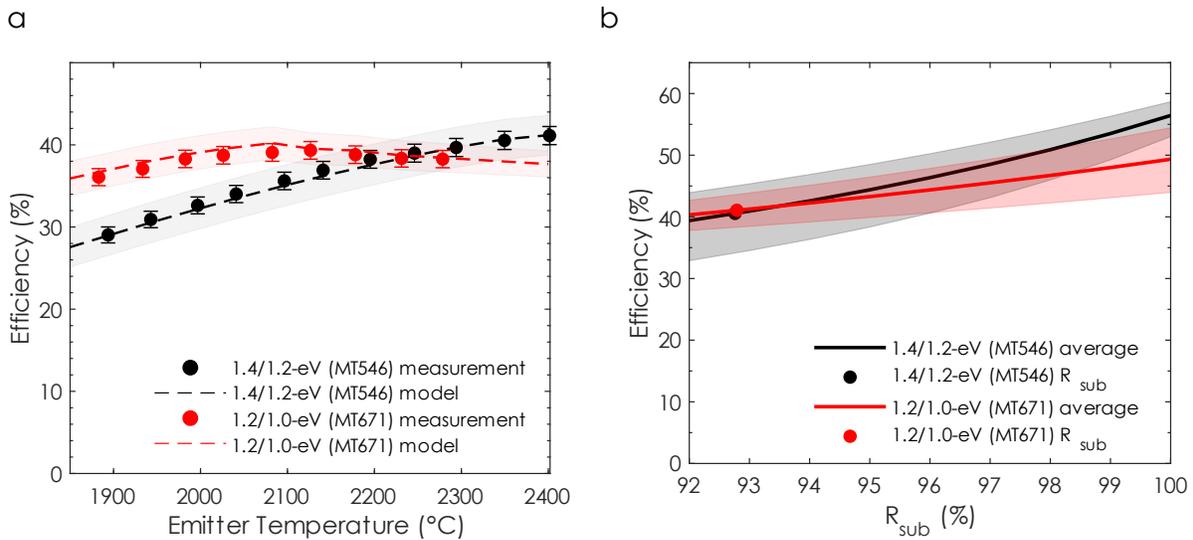

**Figure 3.** a) TPV efficiency measured at different emitter temperatures ranging from ~1900°C to 2400 °C. The dashed lines show the model predictions, and the shaded regions show the uncertainty on the model predictions. b) Predicted efficiency of MT546 and MT671 as weighted sub-bandgap reflectance ($R_{sub}$) is extrapolated assuming a W emitter with AR = 1 and VF = 1 and a 25 °C cell temperature. The solid lines show the average efficiency within the TEGS operating temperature range of 1900°C to 2400°C. The shaded bands show the maximum and minimum efficiency within the temperature range. The dots show the present value of $R_{sub}$ based on the measured reflectance in Fig. 2a weighted by the W AR=1, VF=1 spectrum.

**Conclusions**

We report two-junction TPV cells with efficiencies of > 40% using an emitter with a temperature between ~1900-2400°C. The efficiency of MT546 (1.4/1.2-eV) reaches 41.1±1% at 2400°C, with an average of 36.2% over the target temperature range. The efficiency of MT671 (1.2/1.0-eV) reaches 39.3±1% and varies very little over a wide temperature range with an average efficiency over the ~1900-2300°C temperature range of 38.2%. This new record high performance is enabled by the usage of multi-junction cells with bandgaps ≥ 1.0 eV, which are higher bandgaps than have been traditionally used in TPV. The higher bandgaps enable the use of higher emitter temperatures, which correspond to the temperature range of interest for the low-cost TEGS energy storage technology.[2] This temperature range is also applicable for natural gas or hydrogen combustion, and further demonstrations of integrated systems are warranted.

Reaching 40% efficiency with TPV is remarkable from the standpoint that it now renders TPV as a heat engine technology that can compete with turbines. An efficiency of 40% is already greater than the average turbine-based heat engine efficiency in the USA, but what could make TPV even more attractive than a turbine, is its potential for lower cost (< $0.25/W),[2,26] faster response times, lower maintenance, ease of integration with external heat sources, and fuel flexibility. This is noteworthy because turbine costs and performance have already reached full maturity, so there are limited prospects for future improvement since they are at the end of their development curve. TPV on the other hand is very early in its progress down a fundamentally different development curve. Consequently, TPV has numerous prospects for both improved efficiency (e.g., by improving reflectivity and lowering series resistance) and lowering cost (e.g., by reusing substrates and cheaper feedstocks). Thus, the demonstration of 40% efficiency represents an important step towards realizing the potential which can be achieved with increased attention and funding in the coming years as commercial applications emerge and become profitable.

**Methods**

**TPV Cell Growth and Processing Details**

Fig. 4 shows the device structures of the tandem cells. All materials were grown by atmospheric pressure organometallic vapor phase epitaxy (OMVPE) using trimethylgallium, triethylgallium, trimethylindium, triethylaluminum, dimethylhydrazine, arsine, and phosphine. Diethylzinc and carbon tetrachloride were used as p-type dopants and hydrogen selenide and dislane were used as n-type dopants. Growth took place in a purified hydrogen gas flow of 6 lpm. Substrates were n-type (100) GaAs with a 2° offcut towards the (111)B plane, and all devices were grown in an inverted configuration. For both types of cells, the substrate was prepared by first etching in $NH_4OH : H_2O_2 : H_2O$ (2:1:10 by volume). The substrate was then mounted on a graphite susceptor and heated inductively to 700°C under an arsine overpressure, followed by a ~10 minute deoxidization under arsine.

Growth of the 1.4/1.2-eV tandem started with a 0.2 µm GaAs buffer, then a 0.5 µm GaInP etch stop layer. Then, 0.1 µm of GaInAsN:Se and 0.2 µm of GaAs:Se were deposited as the front contact layer. Then, the top cell was grown, starting with a 0.02 µm AlInP window layer, then a 0.1 GaAs:Se emitter, a 0.1 µm undoped GaAs layer, a 2.8 µm GaAs:Zn base layer, and a 0.12 GaInP back surface field (BSF) layer. Next, a AlGaAs:C/GaAs:Se/AlGaAs:Si quantum well tunnel junction was grown, followed by a GaInP compositionally graded buffer (CGB). The CGB consisted of 0.25 µm GaInP steps spanning the compositional range $Ga_{0.51}In_{0.49}P$ to $Ga_{0.34}In_{0.66}P$ at a rate of 1% strain/μm, with the final layer being a 1 μm $Ga_{0.34}In_{0.66}P$ strain overshoot layer. For the bottom cell, a 1 μm $Ga_{0.37}In_{0.63}P$ window, a 0.1 $Ga_{0.85}In_{0.15}As$:Se emitter, 0.1 $Ga_{0.85}In_{0.15}As$ i-layer, and a 1.5 μm $Ga_{0.85}In_{0.15}As$:Zn base and 0.05 μm $Ga_{0.37}In_{0.63}P$:Zn BSF. Finally, a 0.05 μm $Ga_{0.85}In_{0.15}As$:Zn++ back contact layer was grown.

For the 1.2/1.0-eV design,[43] a 0.2 μm GaAs buffer layer was grown first, then a GaInP CGB consisting of 0.25 μm GaInP steps, spanning the range $Ga_{0.51}In_{0.49}P$ to $Ga_{0.19}In_{0.81}P$, with the final layers being a 1 μm $Ga_{0.19}In_{0.81}P$ strain overshoot layer and a 0.9 μm $Ga_{0.22}In_{0.78}P$ step back layer lattice matched to the in-plane lattice constant of the $Ga_{0.19}In_{0.81}P$. A 0.3 μm $Ga_{0.70}In_{0.30}As$:Se front contact layer was grown next, followed by the top cell, starting with a 0.02 μm $Ga_{0.22}In_{0.78}P$:Se window, a 1 μm $Al_{0.15}Ga_{0.55}In_{0.30}As$:Se emitter, an undoped 0.1 μm $Al_{0.15}Ga_{0.55}In_{0.30}As$ i-layer, a 2.1 μm $Al_{0.15}Ga_{0.55}In_{0.30}As$:Zn base and a 0.07 μm $Ga_{0.22}In_{0.78}P$:Zn BSF. Then the tunnel junction, comprising a 0.2 μm $Al_{0.15}Ga_{0.55}In_{0.30}As$:Zn, a 0.05 μm $GaAs_{0.72}Sb_{0.28}$:C++, and a 0.1 μm $Ga_{0.22}In_{0.78}P$:Se++ layer, was grown. Lastly the bottom cell was grown, comprising a 0.05 μm $Ga_{0.22}In_{0.78}P$:Se window, a 1.5 μm $Ga_{0.70}In_{0.30}As$:Se emitter, a 0.1 μm $Ga_{0.70}In_{0.30}As$:Zn i-layer and 0.02 μm $Ga_{0.22}In_{0.78}P$:Zn BSF. Finally, a 0.05 μm $Al_{0.4}Ga_{0.30}In_{0.30}As$:Zn++ back contact layer was grown.

After growth, a ~2 μm thick reflective gold back contact was electroplated to the exposed back contact layer (the last semiconductor layer grown). The samples were bonded with low viscosity epoxy to a silicon handle and the substrates were etched away in $NH_4OH : H_2O_2$ (1:3 by volume). Gold front grids were electroplated to the front surfaces through a positive photoresist mask, using a thin layer of electroplated nickel as an adhesion layer. The grids were nominally 10 μm wide, 100 μm apart, and ≥ 5 μm thick. The samples were then isolated into individual devices using standard wet-chemical etchants and cleaved into single cell chips for characterization. The completed cells had mesa areas of 0.8075 $cm^2$, with illuminated areas (discounting the single busbar but including the grid fingers) of 0.7145 $cm^2$.

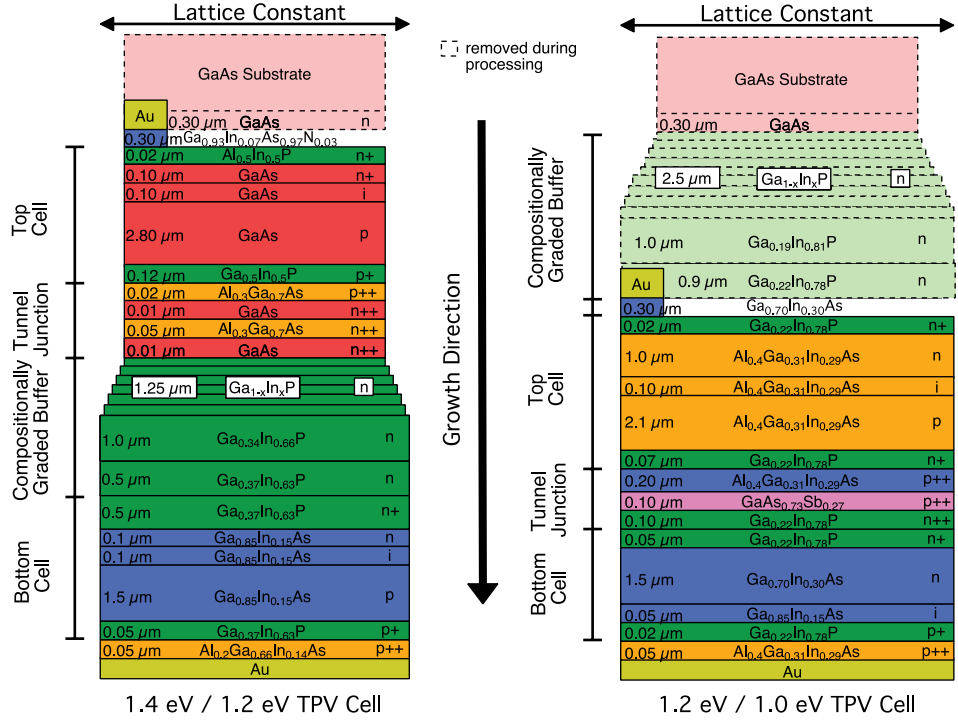

**Figure 4.** Device structures of the 1.4/1.2-eV and the 1.2/1.0-eV tandems.

**Efficiency Measurement**

To measure the TPV cell efficiency, we seek direct measurement of the two contributing quantities in Eq. 1, the power output $P_{out} = V_{oc}I_{sc}FF$ and the heat generated in the cell, $Q_c$. To test the cells under a well-controlled and relevant spectrum (emission from tungsten between 1900-2400°C for TEGS), a tungsten halogen lamp was used in combination with a concentrator (spectra shown in Extended Data Fig. 3). The concentrator consists of a silver-plated elliptical reflector behind the lamp and a compound parabolic reflector (CPC) obtained from Optiforms that further concentrates the light onto the cell. At the base of the CPC, a water-cooled aluminum aperture plate is suspended above the TPV cell (see Fig. 5a and 5b). The area of the aperture is $0.312$ cm$^2$ and the active area of the cell is $0.7145$ cm$^2$.

To keep the TPV cell cool it is mounted on a microchannel copper heat sink (M2, Mikros) that is water-cooled (Fig. 5c and 5d). To measure $Q_{abs}$, a heat flux sensor (HFS), model gSKIN XP obtained from greenTEG, is placed between the cell and the heat sink. Thermally conductive adhesive tape holds the HFS in place on the heat sink, and thermal paste provides thermal contact between the cell and the HFS. Electrical contact to the cell bus bars is accomplished using a pair of copper clips which are both electrically and thermally isolated from the heat sink using a piece of insulation. A pair of wires is connected to the bottom of each copper clip to perform a 4-wire measurement. The bottom side of the aluminum aperture plate is shielded with several layers of copper-coated Kapton and aluminum tape acting as a radiation shield to reduce the radiative transfer between the aperture plate and the TPV cell.

A DC power supply (Magna-Power) provides power to the tungsten halogen lamp and the voltage is controlled to achieve the desired emitter temperature. The lamp is rated for 5 kW at 3200 K, but the temperature and power are tuned down to the desired emitter temperature by controlling the voltage to the lamp using the power supply. The emitter temperature was determined by measuring the resistance of the tungsten heating element in the lamp and using published correlations on the temperature dependence of the electrical resistivity and resistance of tungsten filaments in incandescent lamps.[52] First the cold resistance of the bulb was measured at the point of the bulb junction and at the point of contact with the power supply to determine the resistance of the electrical leads to the bulb. The hot bulb resistance was measured by subtracting the electrical lead resistance from the total resistance as determined from the voltage and current input to the DC power supply. The heat sink was mounted to the z-stage to allow for repeatable control of the TPV cell positioning with respect to the aperture, reflectors, and lamp.

The TPV efficiency was measured by taking simultaneous measurements of $P_{out}$ and $Q_c$. The electric power was measured using a source meter (Keithley 2430) by sourcing the voltage and measuring the current density at the maximum power point, and $Q_c$ was measured using the HFS beneath the cell. Due to the temperature-dependent sensitivity of the HFS the average HFS temperature, $T_s$, is needed, which is taken from the average of the hot and cold side temperatures. The hot side temperature was measured by a thermocouple placed underneath the cell (Extended Data Figure 2). The cold side temperature was determined iteratively using the thermal resistance of the sensor (4.167 K/W), the measured heat flux, and the cell temperature. From the calibration certificate from the manufacturer, the sensitivity $S$ (μV/(W/$cm^2$)) is given by Eq. 2.

$$S = (T_s - 22.5)0.025 + 19.98 \qquad (2)$$

The impact of $T_s$ on $S$ is that $S$ changes by 0.125% per °C temperature change from the reference temperature of 22.5 °C.

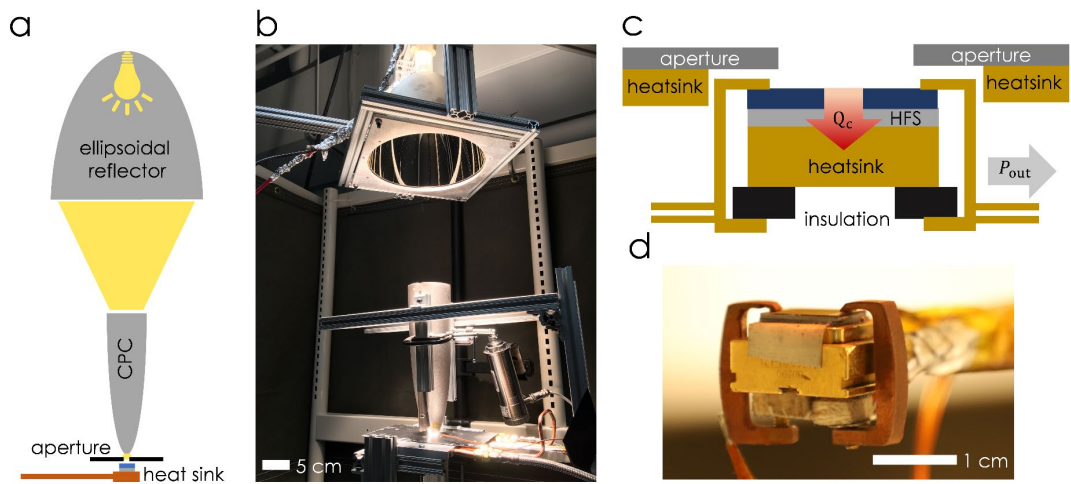

of the concentrator setup. c) Schematic of the heat and electricity flows through the measurement device. Electric power is extracted by two copper clips which interface with the cell bus bars on the top surface of the cell and are thermally and electrically insulated from the heat sink. d) Image of the cell on the heatsink with electrical leads. The aperture was removed for clarity.

**Emitter Spectrum**

The spectrum of the light source was measured using spectrometers in the visible (Ocean Insight FLAME) and in the NIR (Ocean Insight NIRQUEST). The spectrometers were calibrated using a 1000W 3200K quartz tungsten halogen bulb with known spectrum (Newport). Spectrum measurements at several temperatures can be found in Extended Data Fig. 3. To extrapolate the measured spectrum to a broader wavelength range, the spectrum was modeled by considering literature emission of tungsten,[53] the filament material, and transmission of quartz, the envelope surrounding the bulb. Quartz transmission was calculated for a 3 mm thick piece of quartz using optical constants from literature.[54] The filament consists of tungsten coils with non-zero view factor to themselves. The coil geometry acts to smooth the spectral emission because light emitted by the inside of the coil has a high view factor to itself. Therefore, a geometric factor accounting for this smoothing was used as a fitting parameter to model the spectrum to extend it beyond the spectrometer measurement range. Extended Data Fig. 4 shows a comparison between the spectrum described by the emission of tungsten with AR=1 and VF=1, a graybody spectrum shape, and the model which was found to agree well with the measured spectrum (Extended Data Figure 3). Due to the good agreement, the modeled spectrum was then used to form the efficiency predictions. We refer to this spectrum as $E_{TPV}(\lambda, T)$ in the subsequent sections.

Extended Data Fig. 5 shows comparison between the TPV model results under the lightbulb spectra with spectra corresponding to emitters with $VF = 1$, which allows the reflected light to be recycled (an example of these systems is shown in Figure 1a and 1b). Modeling is shown for a tungsten emitter operating with $AR = 1$ and $VF = 1$ and for a blackbody emitter with $VF = 1$. Results show that the lightbulb spectra provide a characterization of TPV efficiency which is relevant to various higher intensity spectra experienced in TPV systems.

**Effective view factor**

To compare the measured TPV cell performance to model predictions, the effective view factor, $VF_{eff}$, was calculated according to the method described by Osterwald[55] and shown in Eq. 3 and Eq. 4. We used an NREL-fabricated GaAs cell with measured EQE and a $J_{sc}$ that was measured at NREL on an XT-10 solar simulator (AM1.5D, 1000 W/m²) using a secondary calibration reference cell to set the intensity. Prior to an efficiency measurement, the GaAs cell was placed in the setup at the same location as the multi-junction cell using the z-stage. In Eq. 3, $J_{sc}^{TPV}$ is the short circuit current of the GaAs cell measured in the efficiency setup, $J_{sc}^{G173d}$ is the short-circuit current of the cell measured using the XT-10 simulator at NREL, $E_{TPV}(\lambda, T)$ is the Planck distribution spectral emissive power under the measured spectrum in the efficiency setup (Extended Data Figure 3), and $E_{G173d}(\lambda)$ is the AM1.5D spectrum. Both spectra are in units of W/m²/nm. We define $VF_{eff}$ as the ratio of the actual irradiance in the efficiency setup, $E_{irradiance}^{TPV}$, to the full irradiance for the Planck distribution spectral emissive power at the same test temperature, $\int E_{TPV}(\lambda, T)d\lambda$, (Eq. 4). The Emitter Spectrum section above discusses how $E_{TPV}(\lambda, T)$ was determined. Measurements of $J_{sc}^{TPV}$ were averaged across the range of emitter temperatures.

$$J_{sc}^{TPV} = VF_{eff} \times \frac{J_{sc}^{G173d}}{1000 \text{ W/m}^2} \times \frac{\int E_{TPV}(\lambda, T) \, EQE(\lambda) \, \lambda \, d\lambda}{\int E_{G173d}(\lambda) \, EQE(\lambda) \, \lambda \, d\lambda} \times \int E_{G173d}(\lambda) d\lambda \quad (3)$$

$$VF_{eff} = \frac{E_{irradiance}^{TPV}}{\int E_{TPV}(\lambda, T)d\lambda} \quad (4)$$

$VF_{eff}$ was then used to form efficiency model predictions. A useful metric to allow for comparisons with other systems is to define an effective view factor in relation to the blackbody spectrum. Eq. 5 compares the TPV irradiance in our efficiency setup to that of the Planck distribution blackbody spectrum at the same test temperature.

$$VF_{eff,black} = \frac{E_{irradiance}^{TPV}}{\int E_B(\lambda, T)d\lambda} \quad (5)$$

Because the shape of $E_{TPV}(\lambda, T)$ varies slightly with temperature, $VF_{eff,black}$ also changes slightly with temperature. Averaged across the emitter temperatures, for MT546 $VF_{eff,black} = 10.07\%$ and for MT671 $VF_{eff,black} = 10.65\%$. The differences are due to slight adjustments made to the setup between measurements of the two multi-junction cells.

**Efficiency Validation**

Eq. 1 for TPV efficiency can also be written in terms of Eq. 6, where $P_{inc}$ is the irradiance incident on the cell, $P_{ref}$ is the flux reflected by the cell, $P_{inc,a}$ is the above-bandgap irradiance, $P_{inc,sub}$ is the sub-bandgap irradiance, $R_a$ is the spectral-weighted above-bandgap reflectance, and $R_{sub}$ is the spectral-weighted sub-bandgap reflectance.[43] The denominator of the efficiency expression represents the net flux to the cell.

$$\eta_{TPV,pairwise} = \frac{P_{out}}{P_{out} + Q_c} = \frac{V_{oc}J_{sc}FF}{P_{inc} - P_{ref}} = \frac{V_{oc}J_{sc}FF}{P_{inc} - P_{inc,a}R_a - P_{inc,sub}R_{sub}} \quad (6)$$

To model the numerator, or electric power portion of the efficiency expression, we used a well-established analytical model which takes values extracted from experiments as input parameters.[56] Using a flash simulator with known spectral irradiance, we first measured the cell performance under carefully controlled conditions of known spectrum and cell temperature fixed at 25°C. Using the model, we fit the data satisfactorily over an irradiance range of several orders of magnitude (shown for MT671 in Extended Data Fig. 6). The fitting was done using only three parameters: the geometric averaged dark current for the two junctions in the form of $E_g/e = W_{oc}$,[57] the n=2 component of the dark current, and the effective lumped series resistance $R_{series}$. We refer to these as the cell characteristic parameters.

We then measured the IV performance parameters ($J_{sc}, V_{oc}, FF$) of the device as a function of the ratio of top to bottom junction photocurrents under a continuous 1 sun simulator for which the spectral content can be varied. Using the measured EQE of the cells (Extended Data Fig. 1), the photocurrent ratio for a given emitter temperature can be calculated, and using reference cells[55] the simulator was set to that photocurrent ratio for each emitter temperature. With the measured EQE and the cell characteristic parameters from above, we calculated the cell performance parameters and compared them to the measurements (shown for MT671 in Extended Data Fig. 7). The agreement supports the validity of the modeling process and its ability to correctly predict performance trends under a wide range of conditions – both irradiance and emitter temperature (i.e. spectrum).

The measured spectra (Extended Data Fig. 3) were used along with the measured EQE to calculate the top and bottom junction photocurrents (Eq. 7). With those as inputs to the model, and the cell characteristic parameters determined above, we computed the cell performance parameters under the actual efficiency measurement conditions. The cell temperature varies (Extended Data Fig. 2). This was accounted for using a well-established model which works especially well for near-ideal devices such as III-V devices. The model accounts for the temperature dependence through its effect on the intrinsic carrier density, and thus the dark current, and the effects of the bandgap variation with temperature.[58,59] Extended Data Fig. 8 shows comparison of the computed cell performance for a 25°C cell and at the measured cell temperature for MT671. Fig. 6 compares the measured $J_{sc}$, $V_{oc}$, and $FF$ with the model predictions.

$$J_{sc} = \frac{q * VF_{eff}}{c * h} \int_0^\infty EQE(\lambda) * E_{TPV}(\lambda, T) \lambda d\lambda \tag{7}$$

The spectral emissive power, $E_{TPV}(\lambda, T)$ was used to determine $P_{inc}$ based on the emitter temperature, $T$, and $VF_{eff}$ (Eq. 8). The reflectance, $\rho(\lambda)$, was measured on two different instruments due to the range of the spectrum. The mid-IR sub-bandgap reflectance was measured using a Fourier-transform infrared (FTIR) spectrometer (Nicolet iS50) with an integrating sphere accessory (PIKE Mid-IR IntegratIR). A copper aperture with area ~0.35 cm² was used over the sample port, and the spot encompassed both the cell and the front grids. The above-bandgap and NIR sub-bandgap reflectance was measured using a UV-Vis-NIR spectrophotometer (Cary 7000) with the diffuse reflectance accessory and with a spot size ~0.4 cm² encompassing the cell and the front grids. $P_{ref}$ was then calculated according to Eq. 9.

$$P_{inc} = VF_{eff} \int_0^\infty E_{TPV}(\lambda, T) d\lambda \tag{8}$$

$$P_{ref} = VF_{eff} \int_0^\infty E_{TPV}(\lambda, T) \rho(\lambda) \, d\lambda \tag{9}$$

This approach to modeling the cells was used to predict the cell performance under the tungsten filament lighting conditions. The decomposition of reflectance into $R_a$ and $R_{sub}$ portions (Eq. 5) enabled the subsequent predictions of efficiency at higher $R_{sub}$ shown in Fig. 3b.

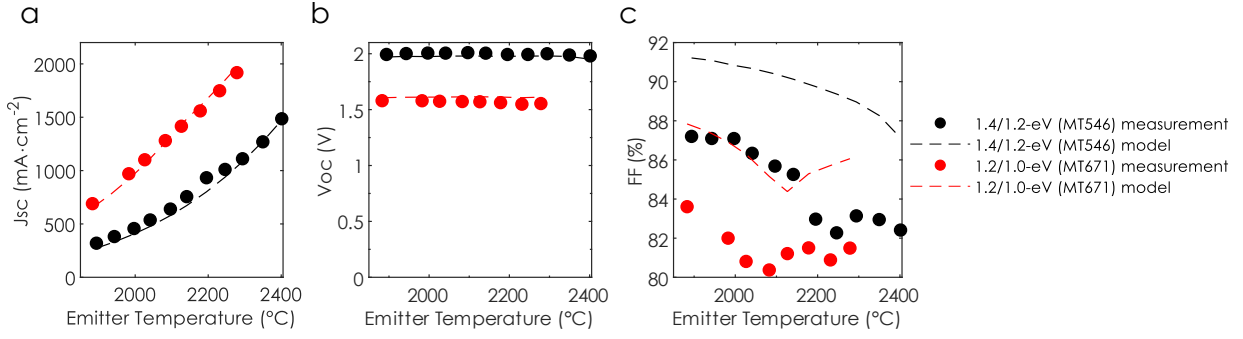

**Figure 6**. Modeled vs measured a) $J_{sc}$, b) $V_{oc}$, and c) $FF$. Good agreement can be seen between the measurement and model predictions. For each device, the $FF$ measurement and model exhibit the same trend and the minimum in $FF$ for MT671 agrees well between the model and measurement which suggests good calibration of the emitter temperature.

**Heat transfer considerations**

We examined the influence of different parasitic heat flows on the efficiency measurement. A schematic of the different parasitic heat flows is shown in Extended Data Fig. 9a and they are quantified in Extended Data Fig. 9b. Possible parasitic heat flows, $Q_{parastic}$, are given by Eq. 10. A positive value of $Q_{parastic}$ would act to increase the measured heat flow and reduce the measured efficiency, while a negative value of $Q_{parastic}$ would have the opposite effect.

$$Q_{parastic} = Q_{cond,clips} + Q_{rad,gain} - Q_{rad,loss} - Q_{conv,loss} \qquad (10)$$

For example, the aperture does not block all the light hitting the electrical leads. $Q_{cond,clips}$ is arises due to conduction from the electric leads into the cell which is cooled by the heat sink, which by design are thermally stranded from the heat sink using insulation. To quantify this value, we performed measurements of the heat flow both with and without the electrical leads attached to the cell. In both cases the cell was operating at $V_{oc}$ to avoid differences in heating due to power being extracted by the cell. The difference between the two heat flows is $Q_{cond,clips}$. Results show that at most emitter temperatures, the heat flow with the presence of the leads is larger than without, because the leads are thermally stranded while the cell is actively cooled. Thus, inclusion of such a term would lead to a higher efficiency than what is reported.

The next parasitic heat flow is due to radiation from the aperture plate to the cell, $Q_{rad,gain}$. The temperature of the bottom of the aperture plate was measured with a thermocouple at the different emitter temperatures. Aperture temperatures varied from 43°C at the lowest emitter temperature to 125°C at the highest. The view factor between the aperture plate and the cell, $F_{ac}$, was calculated from their geometry and spacing. The heat transfer from the aperture to the cell was calculated using a diffuse gray approximation as,

$$Q_{rad,gain} = \frac{\sigma(T_{ap}^4 - T_{cell}^4)}{\frac{1-\varepsilon_{cell}}{\varepsilon_{cell}A_{cell}} + \frac{1}{A_{ap}F_{ac}} + \frac{1-\varepsilon_{ap}}{\varepsilon_{ap}A_{ap}}} \qquad (11)$$

The emissivity of the cell weighted by the spectrum at the aperture temperature is $\varepsilon_{cell}$ (0.15 for MT546 and 0.11 for MT671) and the emissivity of the aperture is $\varepsilon_{ap}$ (~0.1).

There is also radiative transfer between the cell and the ambient environment, $Q_{rad,loss}$, but this was found to be negligible at the cell temperature and the calculated view factor between the cell and the environment. Nonetheless it was included in the calculation of $Q_{parastic}$ for completeness.

Another parasitic heat flow is convective heat loss from the cell to the ambient,

$$Q_{conv,loss} = hA(T_\infty - T_{cell}) \qquad (12)$$

where $h$ is the convective heat transfer coefficient, $A$ is the area of the cell, and $T_\infty$ is the ambient temperature. The ambient temperature was measured with a thermocouple which was blocked from irradiance by the light source using several layers of aluminum foil forming a radiation shield. Ambient temperatures were found to vary between 26°C at the lowest emitter temperature and 33°C at the highest emitter temperature. $h$ was calculated using a Nusselt (Nu) correlation for natural convective heat transfer from a horizontal plate at the calculated Rayleigh (Ra) number.[60] Heat transfer coefficients were calculated at each cell/ambient temperature, with the average being $h = 5.8 \, W/m^2/K$.

Quantifying $Q_{parastic}$ (Extended Data Fig. 9a) shows that $Q_{parastic}$ is a small and positive quantity at most emitter temperatures. At lower emitter temperatures it is dominated by $Q_{cond,clips}$, while at higher emitter temperatures $Q_{conv,loss}$ and $Q_{rad,gain}$ become more important. The potential impact of $Q_{parastic}$ on the efficiency measurement is shown in Extended Data Fig. 10. Overall, $Q_{parastic}$ has a small impact on the efficiency because $Q_{parastic}$ is two orders of magnitude lower than $Q_c$. Because $Q_{parastic}$ is largely derived from modeling and correlation, we do not include it in the efficiency measurement reported. In fact, our calculation of $Q_{parastic}$ largely predicts a higher efficiency than the measured value which suggests the reported measured efficiency could be conservative.

**Uncertainty Propagation**

Uncertainty in the efficiency measurement arises from the measurement of $P_{out}$ and the measurement of $Q_c$ (Eq 1). From the manufacturer, the calibration accuracy of the HFS is ±3%. We include an additional 10°C temperature uncertainty on $T_s$, the sensor temperature, which comes from the average temperature rise across the sensor as calculated from the thermal resistance of the sensor (4.167 K/W) and the average heat flux passing through the sensor. This leads to an uncertainty of heat absorbed of $B_{Qc} = 0.0325 Q_c$. From the source meter, the voltage measurement uncertainty is 0.03% of the voltage ($B_V = (3x10^{-4}) * V$) and the current measurement uncertainty is 0.06% of the current ($B_I = (6x10^{-4}) * I$). This leads to an uncertainty on the electric power measurement of $B_P = \sqrt{(I*B_V)^2 + (V*B_I)^2}$, which is negligible due to the low uncertainty on voltage and current. The absolute uncertainty in measured efficiency, $B_{\eta,measure}$, was calculated as,

$$B_{\eta,measure} = \sqrt{\left(\frac{\partial \eta}{\partial P}B_P\right)^2 + \left(\frac{\partial \eta}{\partial Q_c}B_{Q_c}\right)^2} \qquad (13)$$

The uncertainty in the model prediction primarily arises from the uncertainty in predicted $J_{sc}$ ($B_{Jsc} \approx 0.03 * J_{sc}$) from the uncertainty of the EQE measurement of the multi-junction cell, and from the uncertainty of the FTIR reflectance measurement leading to $B_{R_{sub}} \approx 0.013$. Propagating these errors through Eq. 5, the absolute uncertainty in modeled efficiency, $B_{\eta,model}$, was calculated according to Eq. 14 and the model uncertainty is shown by the shaded regions in Fig. 3a.

$$B_{\eta,model} = \sqrt{\left(\frac{\partial \eta}{\partial J_{sc}}B_{Jsc}\right)^2 + \left(\frac{\partial \eta}{\partial R_{sub}}B_{R_{sub}}\right)^2} \qquad (14)$$

The uncertainty on the emitter temperature measurement was calculated from the variation in resistance of the bulb measured at each emitter temperature and the uncertainty on temperature dependence of resistance from the literature expression that was used which is a 0.1% relative error on resistance as a function of temperature.[52] The RMS of these two yields temperature measurement uncertainties of < 4 °C which had a negligible impact on model uncertainty.

**Extended Data**

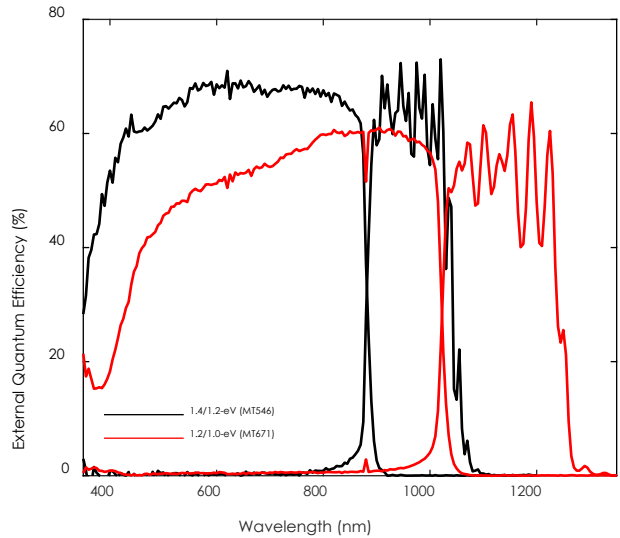

**Extended Data Figure 1.** External quantum efficiency (EQE).

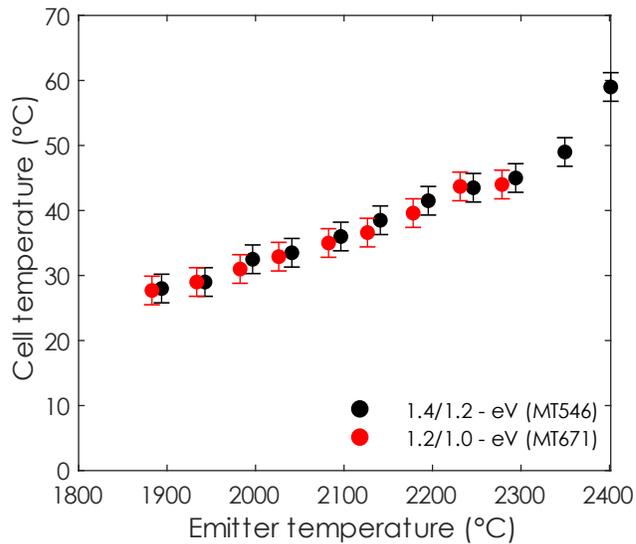

**Extended Data Figure 2**: Cell temperature vs emitter temperature. The cell temperature increases with emitter temperature due to the heat flux sensor which undesirably impedes heat flow.

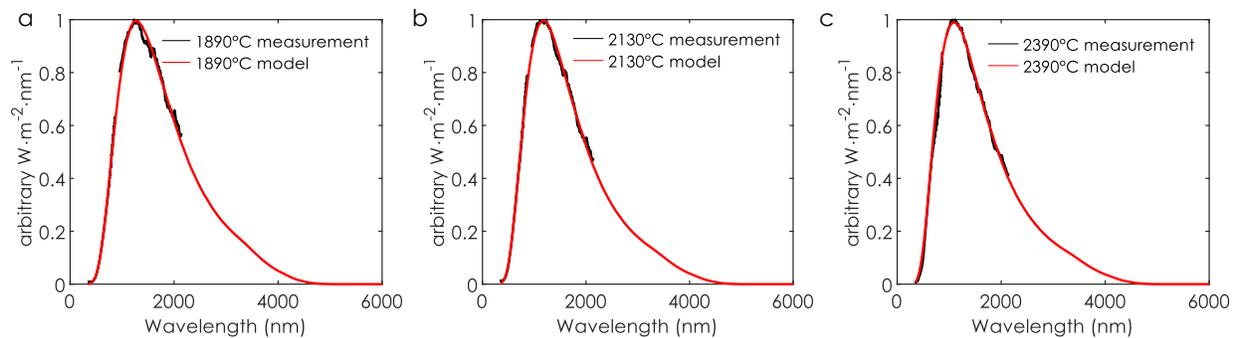

**Extended Data Figure 3**: Emitter spectrum measurements and model used to extend the spectrum across the wavelength range. The spectral radiance goes to zero > ~4500 nm due to the presence of the quartz envelope around the bulb, as quartz is absorbing beyond this wavelength.

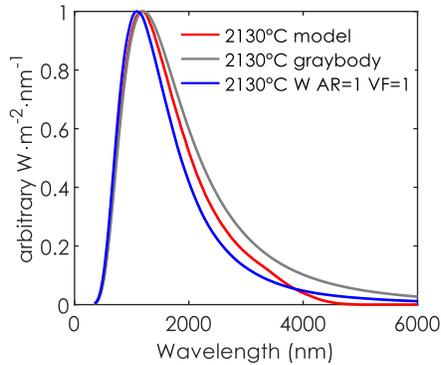

**Extended Data Figure 4**: The red curve shows the modeled spectrum which agrees well with the measurement (see Extended Data Figure 2b). The gray curve shows comparison to a graybody at the same emitter temperature. The blue curve shows comparison to the spectrum described by the literature emission of tungsten with AR=1, VF=1. All curves are normalized by their peak.

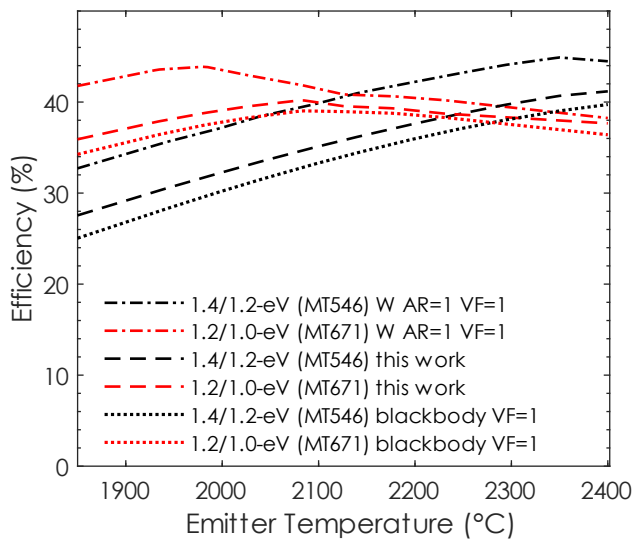

**Extended Data Figure 5**. Comparison of modeled TPV efficiency under the spectrum in this work with emitters which could be incorporated into a TPV system in which the $AR$ and $VF$ allow for the reflected light to be recycled. Shown is a tungsten (W) emitter with $AR = 1$ and $VF = 1$ as well as a blackbody emitter with $VF = 1$. An example of systems which could have this geometry is shown in Figure 1a and 1c. Under these two spectra, the higher irradiance results in higher current density and larger losses due to the series resistance. The W emitter results in a higher efficiency because the selective emissivity properties of W suppress some of the below-bandgap energy. Additionally, the W emitter causes the peak in efficiency to shift to lower temperature because the emissivity of W weights the spectrum towards shorter wavelengths. The blackbody emitter results in a lower efficiency because the high irradiance causes a larger penalty of series resistance loss due to the high current density. The comparison shows that the efficiency measured

under the lightbulb spectrum provides an appropriate and relevant characterization for TPV efficiency in a real TPV system. In all cases, the cell temperature is 25°C.

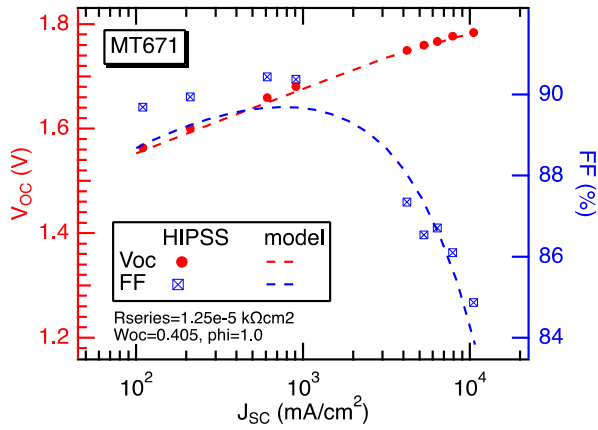

**Extended Data Figure 6.** Measurements of $V_{oc}$ and $FF$ vs $J_{sc}$ for MT671 under the high irradiance flash simulator over a wide range of irradiances, but fixed spectrum and fixed cell temperature at 25°C. A model was fit to the data using the three fitting parameters to determine the cell characteristics. The measurement over a wide irradiance range is critical to extract the $R_{series}$ parameter under the high-irradiance conditions of interest.

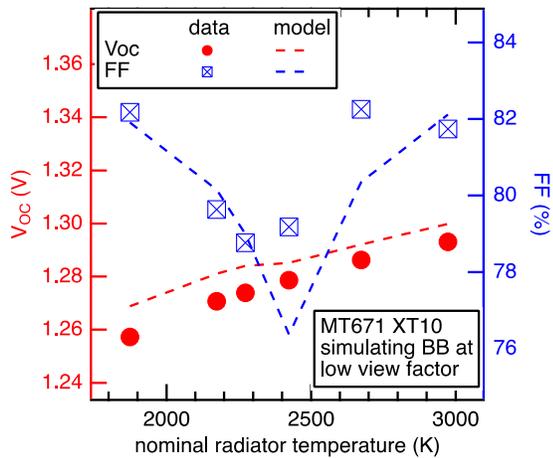

**Extended Data Figure 7.** Low irradiance measurements of $V_{oc}$ and $FF$ under a continuous 1 sun simulator in which the spectral content could be varied to produce photocurrent ratios of the two junctions corresponding to different emitter temperatures. Cell temperature was fixed at 25°C. The model was determined using the cell characteristic parameters which were extracted from fitting to the data over a wide range of irradiances in Extended Data Figure 5. The good agreement suggests that the model can be used to predict $V_{oc}, J_{sc}, FF$ over a wide range of conditions (irradiance and spectra).

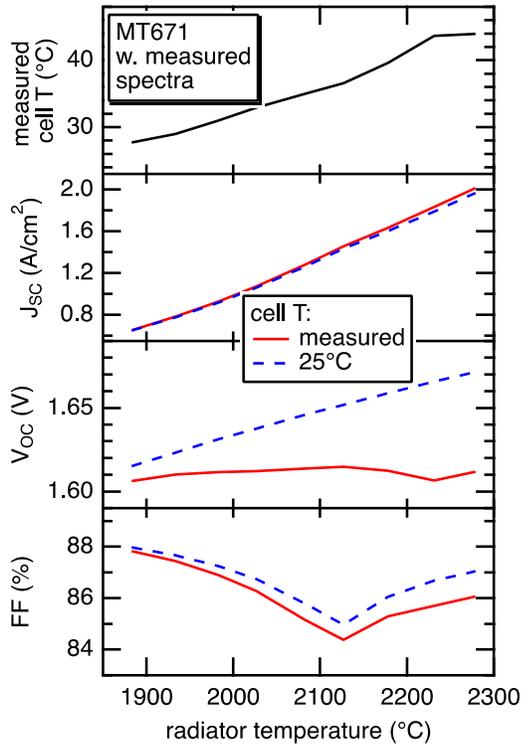

**Extended Data Figure 8.** Modeled cell performance parameters under the measured spectra showing a comparison between results for a 25°C cell temperature and the measured cell temperature.

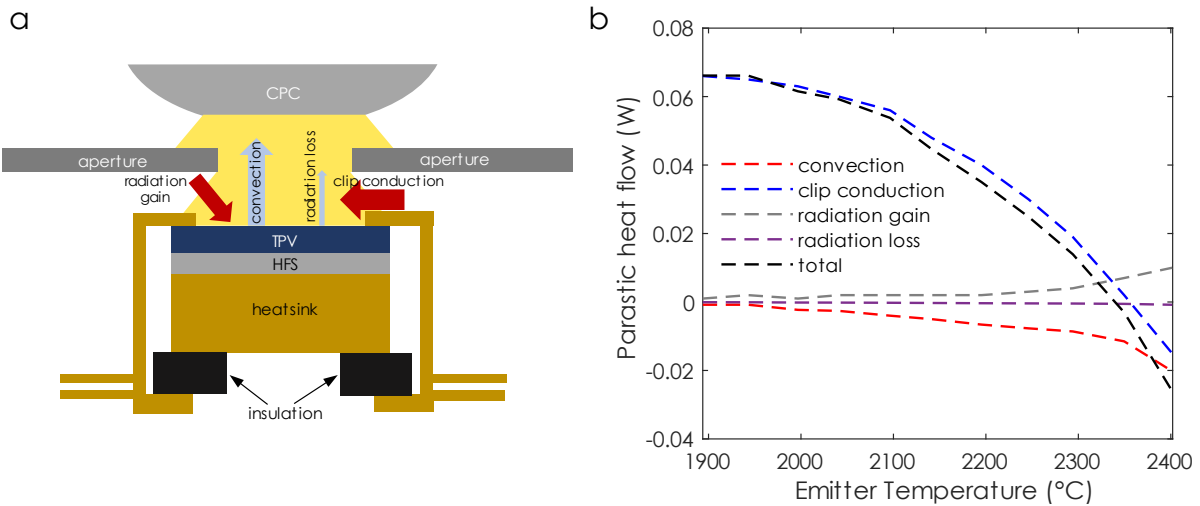

**Extended Data Figure 9.** A) Schematic (not to scale) showing parasitic heat flows in the experiment. B) Calculated parasitic heat flows for MT546

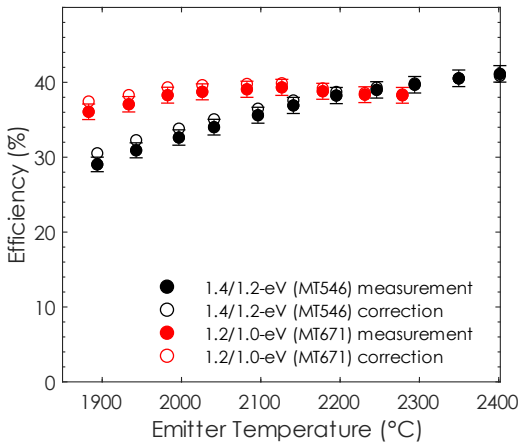

**Extended Data Figure 10.** Comparison of the efficiency measurement (solid circles) to the measurement with the addition of the modeled parasitic heat flows (open circles).

**Extended Data Table 1.** 1.4/1.2 – eV (MT546) measurements of $V_{oc}$, $J_{sc}$, and $FF$.

| Emitter Temperature (°C) | Voc (mV) | Jsc (mA/cm2) | FF (%) |
|---|---|---|---|
| 1894 | 1993.45 ± 2.68 | 319.51 ± 0.19 | 87.20 ± 0.14 |
| 1943 | 2001.50 ± 2.84 | 381.86 ± 0.23 | 87.10 ± 0.15 |
| 1997 | 2005.95 ± 2.58 | 456.95 ± 0.27 | 87.09 ± 0.14 |
| 2041 | 2005.31 ± 2.86 | 537.34 ± 0.32 | 86.34 ± 0.15 |
| 2096 | 2009.89 ± 2.69 | 639.50 ± 0.38 | 85.68 ± 0.14 |
| 2141 | 2005.02 ± 2.86 | 755.95 ± 0.45 | 85.26 ± 0.14 |
| 2195 | 1993.29 ± 2.72 | 933.00 ± 0.56 | 82.97 ± 0.14 |
| 2246 | 1993.43 ± 2.72 | 1010.94 ± 0.61 | 82.27 ± 0.13 |
| 2294 | 2000.00 ± 2.90 | 1111.07 ± 0.67 | 83.14 ± 0.14 |
| 2350 | 1988.69 ± 2.87 | 1269.69 ± 0.76 | 82.95 ± 014 |
| 2401 | 1980.49 ± 2.77 | 1485.82 ± 0.89 | 82.41 ± 0.14 |

**Extended Data Table 2.** 1.2/1.0 – eV (MT671) measurements of $V_{oc}$, $J_{sc}$, and $FF$.

| Temperature (°C) | Voc (mV) | Jsc (mA/cm2) | FF (%) |
|---|---|---|---|
| 1883 | 1579.88 ± 2.13 | 690.12 ± 0.41 | 83.61 ± 0.14 |
| 1934 | 1563.86 ± 2.15 | 804.04 ± 0.48 | 83.35 ± 0.14 |
| 1983 | 1579.35 ± 2.13 | 970.44 ± 0.58 | 82.00 ± 0.13 |
| 2026 | 1574.75 ± 2.13 | 1101.57 ± 0.66 | 80.81 ± 0.13 |
| 2083 | 1572.30 ± 2.15 | 1281.05 ± 0.77 | 80.37 ± 0.13 |
| 2127 | 1570.44 ± 2.15 | 1415.80 ± 0.85 | 81.21 ± 0.13 |
| 2178 | 1562.42 ± 2.13 | 1559.57 ± 0.94 | 81.5 ± 0.13 |
| 2231 | 1548.67 ± 2.17 | 1748.27 ± 1.05 | 80.89 ± 0.13 |
| 2279 | 1554.22 ± 2.13 | 1918.07 ± 1.15 | 81.49 ± 0.13 |

**Data Availability**

The data that support the findings of this study are available from the corresponding author upon reasonable request.

**Acknowledgements**

The authors thank W. Olavarria and A. Kibbler for MOVPE growth work and C. Aldridge for earlier processing work.

The authors thank Timothy McClure at MIT for the usage of FTIR spectroscopy. The authors also thank Yannick Salamin and Reyu Sakakibara at MIT for their assistance in the characterization of the light source.

We acknowledge financial support from the U.S. Department of Energy (DOE): Advanced Research Projects Agency – Energy (ARPA-E) cooperative agreement DE-AR0001005; Office of Energy Efficiency and Renewable Energy grants DE-EE0008381, and DE-EE0008375. This work was authored, in part, by Alliance for Sustainable Energy, LLC, the manager and operator of the National Renewable Energy Laboratory for the U.S. Department of Energy (DOE) under Contract No. DE-AC36-08GO28308. The views expressed in the article do not necessarily represent the views of the DOE or the U.S. Government. The U.S. Government retains and the publisher, by accepting the article for publication, acknowledges that the U.S. Government retains a nonexclusive, paid-up, irrevocable, worldwide license to publish or reproduce the published form of this work, or allow others to do so, for U.S. Government purposes.

A.L. conducted the efficiency measurement experiments and analyzed the data. A.L, K.B., and C.C.K. designed, built, and tested the experimental setup. K.L.S. developed and optimized the epitaxial growth of the cells. D.J.F., M.A.S., and K.L.S designed the cells, and M.A.S. and M.R.Y. fabricated them. D.J.F., M.A.S., and K.L.S along with contributions from R.M.F., E.J.T., A.L, and A.R., characterized and modeled the cells. A.L. and S.V. characterized the light source. A.H.

supervised the work. All authors contributed intellectually to the work's execution and to preparation of the manuscript.

**Competing interest declaration**

M.A.S. and E.J.T. also work on a similar project with a small company that could be perceived as competing.

60	Raithby, G. D. & Hollands, K. G. T. in *Handbook of Heat Transfer* (eds W. M. Rohsenow, J. P. Hartnett, & Y. I. Cho) Ch. 4, (McGraw-Hill, 1998).28